\documentclass[11pt,a4paper]{article}

\usepackage{amssymb}
\usepackage{amsmath}
\usepackage{nccmath}
\usepackage[dvipdfmx]{graphicx}

\usepackage{color}

\newcommand{\di}{\partial}
\newcommand{\tr}{\mathrm{tr}}
\newcommand{\Slash}[1]{{\ooalign{\hfil/\hfil\crcr$#1$}}}
\newcommand{\bra}{\langle}
\newcommand{\ket}{\rangle}
\newcommand{\Bra}{\Bigl\langle}
\newcommand{\Ket}{\Bigr\rangle}
\newcommand{\dx}{{d^4}\kern-.07cm{x}}

\newcommand{\Hvs}{\tr[\bar{H}_v H_v]}

\newcommand{\MeV}{\mathrm{\ MeV}}

\newcommand{\Lhat}{\hat L}

\begin{document}

\begin{titlepage}
\begin{center}

\vspace*{10mm}

{\LARGE\bf
An approach to the instanton effect in B system}

\vspace*{20mm}

{\large
Noriaki Kitazawa\footnote[1]{noriaki.kitazawa@tmu.ac.jp}
 and Yuki Sakai\footnote[2]{sakai-yuki1@ed.tmu.ac.jp}
 }
\vspace{6mm}  

{\it
Department of Physics, Tokyo Metropolitan University,\\
Hachioji, Tokyo 192-0397, Japan\\
}

\vspace*{15mm}

\begin{abstract}

We discuss the constraint on the size of QCD instanton effects in low-energy effective theory.
Among various instanton effects in meson mass spectrum and dynamics,
 we concentrate on the instanton-induced masses of light quarks.
The famous instanton-induced six-quark interaction, so-called 't Hooft vertex,
 could give non-perturbative quantum corrections to light quark masses.
Many works have already been achieved to constrain the mass corrections
 in light meson system, or the system of $\pi$, $K$, $\eta$ and $\eta'$,
 and now we know for a fact that the instanton-induced mass of up-quark is too small
 to realize the solution of the strong CP problem by vanishing current mass of up-quark.
In this work
 we give a constraint on the instanton-induced mass correction to light quarks
 from the mass spectrum of heavy mesons, $B^+$, $B^0$, $B_s$ and their anti-particles.
To accomplish this,
 the complete second order chiral symmetry breaking terms are identified
 in heavy meson effective theory.
We find that
 the strength of the constraint from heavy meson masses is at the same level of that from light mesons,
 and it would be made even stronger by more precise data from future B factories and lattice calculations.

\end{abstract}

\end{center}
\end{titlepage}

\section{Introduction}

The instanton effect is a non-perturbative effect
 which is described by non-trivial gauge field configurations in Euclidean space-time \cite{Belavin:1975fg}.
In quantum chromodynamics (QCD)
 the light mesons, $\pi$, $K$ and $\eta$, are identified as pseudo Nambu-Goldstone bosons
 of spontaneous chiral symmetry breaking and obtain their masses by non-zero quark mass effect.
However, the physical $\eta'$ mass is too heavy to be identified
 as a pseudo Nambu-Goldstone boson of U(1)$_A$ symmetry breaking.
It is expected that the instanton effect gives a solution to this problem, the so-called U(1)$_A$ problem \cite{Weinberg:1975ui,tHooft:1986ooh}.
But, there are also some indications of that the instanton effect may not provide a solution to this problem.
The problem could be understood within the $1/N_c$ expansion
 \cite{Witten:1978bc,Witten:1979vv,Veneziano:1979ec}.
There is also an indication of inconsistency between Ward-Takahashi identity for U(1)$_A$ current and quark condensation
 in the instanton configuration \cite{Crewther:1977ce}.
In addition, the instanton effect has not been directly confirmed by experiments yet.

If we believe the existence of the instanton effect, the non-trivial vacuum structure supplies the so-called $\Theta$-term
 which gives rise to CP violation in QCD.
But the $\Theta$-term should be strongly suppressed by some reasons.
This is called the strong CP problem.
The Peccei-Quinn mechanism \cite{Peccei:1977hh} is a possible way to solve this problem.
The predicted new particle, called the axion, 
 which is a Nambu-Goldstone boson with the spontaneous breaking of Peccei-Quinn symmetry,
 has not been discovered yet.
The verification of the instanton effect in the real world remains to be achieved.
 
It is highly important to directly observe instanton-induced effects in experiments.
The instanton effect gives rise to a six-quark interaction,
 which violates U(1)$_A$ symmetry in QCD, known as the 't~Hooft vertex \cite{tHooft:1976snw}.
The contribution of instanton effects in deep inelastic scattering is investigated with instanton perturbation theory \cite{Ringwald:1998ek}
and direct searches have been done at the electron-proton collider HERA \cite{Chekanov:2003ww,Adloff:2002ph,H1:2016jnv}.
No signal is observed, which gives a constraint on the cross section for instanton-induced processes.
This is one of the quantitative result of the direct search for instanton effect.

The six-quark interaction also induces a quantum correction to light quark masses.
This quantum correction is proportional to the product of different quark flavor masses.
An ``effective up-quark mass" of this form was first considered in connection with the instanton effect
 in \cite{Caldi:1977rc,Georgi:1981be}.
Since the instanton effect could generate a non-zero effective up-quark mass even when $m_u = 0$,
the strong CP phase could be unphysical,
 and there would be no strong CP problem\footnote{Here, $m_u$ is a bare or current quark mass, and $m_u=0$ means the existentce of chiral symmetry for up quark.}.
A hidden symmetry of the so-called instanton transformation,
 which is related to the instanton-induced quark mass correction,
 was discovered in low-energy light meson effective theory with next-to-leading order terms in the chiral Lagrangian \cite{Kaplan:1986ru}.
The instanton effect on the second order coupling constant
 has been argued in \cite{Choi:1991nq}.
This attempt is one of the other quantitative results related to the indirect search for instanton effect in light meson system.

The precise data on B meson masses, namely $b$-flavored pseudoscalar mesons $B^+$, $B^0$, $B_s$ and their anti-particles,
 are obtained by various experiments.
The purpose of this article is to give another quantitative result for heavy meson systems.
We treat heavy meson systems using heavy meson effective theory.
In order to discuss the constraint on the size of the instanton effect,
the chiral symmetry breaking terms in the next-to-leading order effective Lagrangian are systematically investigated.
We find a hidden symmetry under instanton transformation in the heavy meson effective theory.
We estimate the possible maximum of correction to light quark masses from the instanton-induced effect under some assumptions 
 and also discuss whether or not the instanton-induced effective mass is large enough to resolve strong CP problem by making $m_u = 0$.

This paper is organized as follows.
In Sec. \ref{sec:LMET}, the effective theory
 which describes the interactions of pseudo Nambu-Goldstone bosons at low energies is introduced.
The dynamical meaning of instanton transformation, which is related to instanton-induced mass correction,
 is discussed.
The light meson mass formulae of next-to-leading order in chiral expansion are derived.
We extract the value of couplings which are sensitive to the instanton effect using the formulae,
 and the constraint on the quark mass correction given by the instanton effect is discussed. 
In Sec. 3, we discuss the instanton effect in the heavy meson effective theory.
The effective Lagrangian which includes the next-to-leading order of chiral symmetry breaking terms
 is constructed.
We show the invariance under the instanton transformation even in heavy meson effective theory.
The mass formulae for pseudoscalar B mesons and 
 the formulae for their mass differences are given.
The constraint on the instanton-induced effect are obtained for the B meson system.
In Sec. 4, we summarize and give some comments.

\section{Constraints in light meson system}
\label{sec:LMET}

We start with the conventional SU(3)$_L\times$SU(3)$_R\times$U(1)$_A$ chiral effective Lagrangian.
The interactions of Nambu-Goldstone bosons at low energies are described by the effective Lagrangian.
We define the effective fields
	\begin{eqnarray}
	\Sigma = U \exp\left( \frac{2 i }{\sqrt{6} f'} \eta' \right) ,\quad U = \exp\left( \frac{2i}{f} \Pi \right),
	\end{eqnarray}
where $f$ and $f'$ are parameters of dimension one.
The field $U$ is transformed as
\begin{eqnarray}
U \rightarrow V_L U V_R^\dagger
\end{eqnarray}
 under SU(3)$_L\times$SU(3)$_R\times$U(1)$_A$ chiral transformation,
 where $V_L$ and $V_R$ are elements of SU(3)$_L$ and SU(3)$_R$, respectively.
The field $U$ consists of more fundamental field
\begin{eqnarray}
\xi = e^{i\Pi/f}
\end{eqnarray}
which transforms non-linearly under the chiral transformation as (see ref.~\cite{Bando:1987br} for review) 
\begin{eqnarray}
\xi \rightarrow V_L \xi  h(\Pi,V_L,V_R)^\dagger = h(\Pi,V_L,V_R) \xi V_R^\dagger.
\end{eqnarray}
Here, the unitary matrix $h(\Pi,V_L,V_R)$ has a property
 that $h(\Pi,V_L,V_R) = V$ under the SU(3) vector transformation of $V_L = V_R = V$.
The field $\Sigma$ is also transformed under SU(3)$_L\times$SU(3)$_R\times$U(1)$_A$ chiral transformation as 
\begin{eqnarray}
\Sigma \rightarrow e^{i\theta_A} V_L \Sigma V_R^\dagger,
\end{eqnarray}
where $e^{i\theta_A}$ is an element of U(1)$_A$.

The field $\Pi$ includes the Nambu-Goldstone bosons as
	\begin{eqnarray}
	\Pi = \frac{1}{\sqrt{2}} 
		\begin{pmatrix}
			\cfrac{\pi^0}{\sqrt{2}} + \cfrac{\eta}{\sqrt{6}} & \pi^+ & K^+ \\
			\pi^- & -\cfrac{\pi^0}{\sqrt{2}} + \cfrac{\eta}{\sqrt{6}} & K^0 \\
			K^- & \overline{K}^0 & - \cfrac{2}{\sqrt{6}}\eta 
		\end{pmatrix}.
	\end{eqnarray}
Since $\Pi^\dagger = \Pi$, $U$ and $\Sigma$ are unitary.
The effective Lagrangian consists of all the SU(3)$_L\times$SU(3)$_R\times$U(1)$_A$ symmetric terms
 described by $\Sigma$ and its derivatives.
Since we consider Nambu-Goldstone bosons at low energies,
 the Lagrangian is expanded in powers of derivatives.
The first order term, which includes only two derivatives, namely an ${\cal O}(p^2)$ term, is
\begin{eqnarray}
{\cal L} = \frac{f^2}{4}\tr (\di_\mu \Sigma \di^\mu \Sigma).
\end{eqnarray}

We perturbatively include the effects of chiral symmetry breaking by quark masses.
Introduce
	\begin{eqnarray}
	\chi = 2 B_0 {\cal M} , \quad {\cal M} = \begin{pmatrix} m_u & 0 & 0 \\ 0 & m_d & 0 \\ 0 & 0 & m_s \end{pmatrix},
	\end{eqnarray}
where $B_0$ is a constant of mass dimension one, 
which is related to the quark condensation.
The QCD Lagrangian is invariant under SU(3)$_L\times$SU(3)$_R\times$U(1)$_A$ chiral transformation,
 if the mass matrix ${\cal M}$ transforms appropriately.
We assume the same symmetry in the effective Lagrangian under the transformation
\begin{eqnarray}
	\label{eq:chitransformation}
\chi \rightarrow e^{i\theta_A} V_L \chi V_R^\dagger.
\end{eqnarray}
This $\chi$ is considered as a quantity of ${\cal O}(p^2)$ in the effective Lagrangian.
Since ${\cal M}$ is just an expansion parameter in the chiral perturbation theory, we can use
\begin{eqnarray}
	\label{eq:massredefinition}
	{\cal M}^{\mathrm{eff}} =  {\cal M} + \frac{2B_0\omega}{( 4 \pi f)^2} (\det {\cal M}^\dagger)({\cal M}^\dagger)^{-1}
\end{eqnarray}
instead of the original ${\cal M}$,
 where ${\cal M}^{\mathrm{eff}}$ has the same transformation property of ${\cal M}$.

The ${\cal O}(p^4)$ effective Lagrangian with quark masses is
	\begin{eqnarray}
	\label{eq:LMEL}
	{\cal L}_{\mathrm{chi}} &=& \frac{f^2}{4} \Bra \di_\mu \Sigma^\dagger \di^\mu \Sigma \Ket  + \frac{f^2}{4} \Bra \chi \Sigma^\dagger + \chi^\dagger \Sigma \Ket \nonumber \\[5pt]
	&&+ L_1 \Bra \di_\mu \Sigma^\dagger \di^\mu \Sigma \Ket ^2 + L_2 \Bra \di_\mu \Sigma^\dagger \di_\nu \Sigma \Ket \Bra \di^\mu \Sigma^\dagger \di^\nu \Sigma \Ket \nonumber \\[5pt]
	&&+ L_3 \Bra \di_\mu \Sigma^\dagger \di^\mu \Sigma \di_\nu \Sigma^\dagger \di^\nu \Sigma \Ket + L_4 \Bra \di_\mu \Sigma^\dagger \di^\mu \Sigma \Ket \Bra \chi \Sigma^\dagger + \chi^\dagger \Sigma \Ket \nonumber \\[5pt]
	&&+ L_5 \Bra \di_\mu \Sigma^\dagger \di^\mu \Sigma ( \chi \Sigma^\dagger + \chi^\dagger \Sigma) \Ket + L_6 \Bra \chi \Sigma^\dagger + \chi^\dagger \Sigma \Ket^2 \nonumber \\[5pt]
	&&+ L_7 \Bra \chi \Sigma^\dagger - \chi^\dagger \Sigma \Ket^2 + L_8 \Bra \chi \Sigma^\dagger \chi \Sigma^\dagger + \chi^\dagger \Sigma \chi^\dagger \Sigma \Ket ,
	\end{eqnarray}
where $\bra A \ket$ denotes the trace of the matrix $A$ over light flavor indices
 and the low-energy coupling constants, $L_i$, are dimensionless parameters \cite{Gasser:1984gg}.

\begin{figure}[t]
	\begin{center}
		\includegraphics[width=6cm]{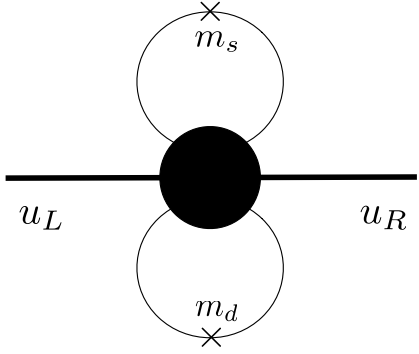}
		\caption{The instanton mass correction to up-quark mass by the 't~Hooft vertex.}
		\label{fig:instantoncorr}
	\end{center}
\end{figure}

The chiral effective Lagrangian has the following symmetries
 which have well-defined dynamical meaning in QCD.
The Lagrangian of eq.(\ref{eq:LMEL}) is invariant under
\begin{eqnarray}
\label{eq:multplicativesym}
	{\cal M} \rightarrow \gamma {\cal M} &,&\ B_0 \rightarrow \gamma^{-1}B_0,
\end{eqnarray}
where $\gamma$ is a multiplicative renormalization factor.
This symmetry is the consequence of the fact
 that the physics is independent of the scale of multiplicative renormalization of quark masses ${\cal M}$.
The Lagrangian of eq.(\ref{eq:LMEL}) is also invariant under the transformation, which is known as the instanton transformation
\begin{eqnarray}
\label{eq:instantontransformation}
	\chi &\rightarrow& \chi + \frac{\omega}{(4\pi f)^2} (\det \chi^\dagger )(\chi^\dagger)^{-1}, 
\end{eqnarray}
and
\begin{equation}
	\label{eq:ITofLC}
	\begin{split}
	L_6 &\rightarrow  L_6 - \frac{\omega}{(16\pi)^2}, \\
	L_7 &\rightarrow  L_7 - \frac{\omega}{(16\pi)^2}, \\
	L_8 &\rightarrow  L_8 + 2\frac{\omega}{(16\pi)^2}.
	\end{split}
\end{equation}
Note that the parameter $\omega$ is invariant under the multiplicative renormalization of quark masses.
It is easy to check the invariance using the following matrix identity
\begin{eqnarray}
	\label{eq:matrixidentity}
	\det M = M^3 - M^2 \tr M - \frac{1}{2} M [ \tr(M^2) - (\tr M)^2],
\end{eqnarray}
where $M$ is any 3$\times$3 complex matrix.
This symmetry is related to the instanton-induced quark mass corrections.
The instanton-induced mass is proportional to the product of different quark flavor masses (see Fig.\ref{fig:instantoncorr}).
The instanton effect gives a six-quark interaction, known as the 't~Hooft vertex,
which induces corrections to the light quark masses.
The physics should be independent of whether or not the instanton correction is included in $\chi$ or $L_i$
 \cite{Georgi:1981be,Choi:1991nq,Choi:1988sy}.
Note that the couplings $L_6,$ $L_7$ and $L_8$ are transformed under the instanton transformation.
We do not pay attention to $L_6$
 since we can not extract it by meson masses 
 (we shortly see that $L_6$ enters in the same form in each mass formula).
We can expect that particularly $L_7$,
 which gives the contribution of the type in Fig.\ref{fig:instantoncorr} in meson mass formulae,
 is produced dominantly by the instanton dynamics \cite{Leutwyler:1989pn,Choi:1991nq},
 though the other couplings, $L_6$ and $L_8$, should be also sensitive to the instanton effect. 
We estimate the value of the coupling $L_7$ in the following.

%
%
%
%
%
%
%
%

From the effective Lagrangian (\ref{eq:LMEL}), we have the meson mass formulae of ${\cal O}(p^4)$:
\begin{eqnarray}
m_{\pi^0}^2 &=& B_0(m_u + m_d) \Biggl[ 1 + \left( \Lhat_6 -\frac{\Lhat_4}{2}\right)(m_u + m_d + m_s) - \frac{\Lhat_5}{4} (m_u + m_d) \nonumber \\
&&\qquad \qquad \qquad \quad \ + \Lhat_7 \frac{(m_u - m_d)^2}{m_u + m_d} + \Lhat_8 \frac{m_u^2 + m_d^2}{m_u + m_d} \Biggr],
\end{eqnarray}
\begin{eqnarray}
m_{\pi^{\pm}}^2 = B_0(m_u + m_d) \Biggl[ 1 + \left( \Lhat_6 -\frac{\Lhat_4}{2}\right)(m_u + m_d + m_s) + \left( \frac{\Lhat_8}{2} - \frac{\Lhat_5}{4} \right)(m_u + m_d) \Biggr],
\end{eqnarray}
\begin{eqnarray}
m_{K^{0}}^2 = B_0(m_d + m_s) \Biggl[ 1 + \left( \Lhat_6 -\frac{\Lhat_4}{2} \right)(m_u + m_d + m_s) + \left( \frac{\Lhat_8}{2} - \frac{\Lhat_5}{4} \right)(m_d + m_s) \Biggr], 
\end{eqnarray}
\begin{eqnarray}
m_{K^{\pm}}^2 = B_0(m_u + m_s) \Biggl[ 1 + \left( \Lhat_6 -\frac{\Lhat_4}{2} \right)(m_u + m_d + m_s) + \left( \frac{\Lhat_8}{2} - \frac{\Lhat_5}{4} \right)(m_u + m_s) \Biggr], 
\end{eqnarray}
\begin{eqnarray}
m_{\eta}^2 &=& B_0 \frac{m_u + m_d + 4m_s}{3} \Biggl[ 1 + \left( \Lhat_6 -\frac{\Lhat_4}{2} \right)(m_u + m_d + m_s) -\frac{\Lhat_5}{12} (m_u + m_d + 4m_s) \nonumber \\
&&\qquad \qquad \qquad \qquad \quad \ + \Lhat_7 \frac{( m_u + m_d - 2m_s )^2}{m_u + m_d + 4m_s} + \Lhat_8 \frac{m_u^2 + m_d^2 + 4m_s^2}{m_u + m_d + 4m_s} \Biggr], 
\end{eqnarray}
\begin{eqnarray}
m_{\eta'}^2 &=& \frac{2}{3}\frac{f^2}{f'^2} B_0 (m_u + m_d + m_s) \Biggl[ 1 + \left( \Lhat_7 + \Lhat_6 - \frac{\Lhat_5}{6} - \frac{\Lhat_4}{2} \right) (m_u + m_d + m_s) \nonumber \\
&&\qquad \qquad \qquad \qquad \qquad \qquad + \Lhat_8 \frac{m_u^2 + m_d^2 + m_s^2}{m_u + m_d + m_s} \Biggr],
\end{eqnarray}
where $\Lhat_i \equiv 32B_0 L_i/f^2$ and we neglect mass mixing between neutral mesons, $\pi^0$, $\eta$ and $\eta'$.
The quantum electromagnetic dynamics (QED) correction to the mass squared of the meson $P$ is proportional to the square of its charge $Q_P$ as  \cite{Dashen:1969eg}
\begin{eqnarray}
\hat{m}^2_P = m^2_P + e^2Q_P^2 C,
\end{eqnarray}
where $C$ is a constant and $\hat{m}$ means the observable mass, which is measured by experiments.
If the QED correction is turned off,
 $\pi^+$ and $\pi^0$ become degenerate in leading order, ${\cal O}(p^2)$.
Therefore the quantity $e^2C$ can be determined in good approximation
 from the observed value as $e^2C \equiv \hat{m}^2_{\pi^\pm} - \hat{m}^2_{\pi^0}$.

To obtain the values of light quark masses at the energy scale where the effective theory is applicable, 
we fit the light quark masses with the mass formulae of mesons {\it in leading order}
 including QED corrections
 and with the value of $B_0$ given by lattice calculations under the non-perturbative RU-MOM renormalization scheme in \cite{Allton:2008pn}\footnote{The quark masses are usually given in $\overline{\mathrm{MS}}$ renormalization scheme. In our present work we do not need to take $\overline{\mathrm{MS}}$ scheme, since we do not use the values in eq.({\ref{eq:fittedmasses}}) to compare with other determinations of the quark masses.}:
\begin{equation}
\label{eq:fittedmasses}
	\begin{split}
m_u &= 2.78 \pm 0.19 \MeV,  \\
m_d &= 4.97 \pm 0.34 \MeV, \\
m_s &= 100.4 \pm 6.8 \MeV. 
	\end{split}
\end{equation}
These results are consistent with the quark mass ratio $m_u/m_d$ given by lattice calculation in \cite{Horsley:2015eaa}.
The parameters are fitted order by order in a spirit of chiral expansion theory.
Once the light quark masses are determined at leading order, the freedom of instanton transformation is fixed,
 since the effective theory up to leading order does not have an invariance under the instanton transformation.
The instanton corrections are included in the next-to-leading order terms.

Now we are going to determine the values of $L_5$, $L_8$, and $L_7$ in order
 by using observable masses and decay constants.
The quantity $L_5/f^2$ can be fixed by using formulae of decay constants in next-to-leading order, ${\cal O}(p^4)$, as
\begin{eqnarray}
\frac{L_5}{f^2} = \frac{f_{K^\pm} - f_{\pi^\pm}}{f_{\pi^\pm}} \frac{1}{4(\hat{m}_{K^{\pm}}^2 - \hat{m}_{\pi^{\pm}}^2 )},
\end{eqnarray}
where
\begin{eqnarray}
f_{\pi^{\pm}} = f \left[ 1 + \frac{8B_0L_4}{f^2} (m_u + m_d + m_s) + \frac{4B_0L_5}{f^2} (m_u + m_d) \right], \nonumber \\
f_{K^\pm} = f \left[ 1 + \frac{8B_0L_4}{f^2} (m_u + m_d + m_s) + \frac{4B_0L_5}{f^2} (m_u + m_s) \right].
\end{eqnarray}
The experimental values
\begin{eqnarray}
f_{\pi^\pm} &=& 92.4 \pm 0.2 \MeV, \nonumber \\
f_{K^\pm} &=&  113.0 \pm 1.0 \MeV,
\end{eqnarray}
 are obtained from the decay processes $\pi^+ \rightarrow \mu^+ \nu_\mu,\  \mu^+ \nu_\mu\gamma$
  and $K^+ \rightarrow \mu^+ \nu_\mu,\ \mu^+ \nu_\mu\gamma$, respectively \cite{Olive:2016xmw},
 and we obtain 
\begin{eqnarray}
\frac{L_5}{f^2} = (2.5 \pm 0.1)\times 10^{-7} \MeV^{-2}.
\end{eqnarray}
We can determine $L_8/f^2$ from the ${\cal O}(p^4)$ relation
\begin{eqnarray}
\frac{m_{K^{0}}^2 - m_{K^{\pm}}^2}{m_{\pi^{\pm}}^2} &=& \frac{ \hat{m}_{K^{0}}^2 - \hat{m}_{K^{\pm}}^2 + \hat{m}_{\pi^{\pm}}^2 - \hat{m}_{\pi^{0}}^2 }{\hat{m}_{\pi^{0}}^2 } \nonumber \\
&=& \frac{m_d - m_u}{m_d + m_u} \left[ 1 + \left( \frac{16L_8}{f^2} - \frac{8L_5}{f^2} \right) (\hat{m}_{K^{\pm}}^2 + \hat{m}_{K^{0}}^2 - \hat{m}_{\pi^{\pm}}^2 ) \right]
\end{eqnarray}
as
\begin{eqnarray}
\frac{L_8}{f^2} = (1.24 \pm 0.06)\times 10^{-7} \MeV^{-2},
\end{eqnarray}
with the quantity
\begin{eqnarray}
R \equiv \frac{m_d + m_u}{m_d -m_u} = 3.53 \pm 0.01
\end{eqnarray}
determined at leading order.
At next-to-leading order, we can derive the relation
\begin{eqnarray}
\frac{2 m_{K^{\pm}}^2 + 2 m_{K^{0}}^2 - m_{\pi^{\pm}}^2 -3 m_\eta^2}{m_\eta^2 - m_{\pi^{0}}^2 } &=& \frac{2 \hat{m}_{K^{\pm}}^2 + 2 \hat{m}_{K^{0}}^2 - 2 \hat{m}_{\pi^{\pm}}^2  + \hat{m}_{\pi^{0}}^2 - 3 \hat{m}_\eta^2}{\hat{m}_\eta^2 - \hat{m}_{\pi^{0}}^2 } \nonumber \\
&=& \frac{8}{f^2} ( L_5 + 6L_7 + 3L_8)(3 \hat{m}_{\pi^{0}}^2 - \hat{m}_{K^{\pm}}^2 - \hat{m}_{\pi^{0}}^2 - \hat{m}_{K^{0}}^2 ) \nonumber \\
&&- \frac{48}{f^2}L_5 \frac{ (\hat{m}_{\pi^{0}}^2 - \hat{m}_{K^{0}}^2 )(\hat{m}_{\pi^{\pm}}^2 - \hat{m}_{K^{\pm}}^2) }{3 \hat{m}_{\pi^{0}}^2 - \hat{m}_{K^{\pm}}^2 - \hat{m}_{\pi^{0}}^2 - \hat{m}_{K^{0}}^2}
\end{eqnarray}
neglecting mass mixing of neutral mesons and we finally obtain
\begin{eqnarray}
\label{eq:L_7}
\frac{L_7}{f^2} = ( -5.1 \pm 0.3 )\times 10^{-8} \MeV^{-2}.
\end{eqnarray}

Assuming that all the value of $L_7$ is produced by the instanton dynamics, we have
 \begin{eqnarray}
 \omega_\mathrm{max} = 0.4 \pm 0.1,
 \end{eqnarray}
using eq.(\ref{eq:ITofLC}) with the result of lattice calculation of $f = 54.1 \pm 4.0$ MeV \cite{Allton:2008pn}.
This conservative maximum value of $\omega$ cannot reproduce the situation $m_u = 0$
 with a sufficiently large value of $m^\mathrm{eff}_u$ in ${\cal M}^\mathrm{eff}$ in eq.(\ref{eq:massredefinition}),
 and, thus, be a solution of strong CP problem.
In fact, the instanton effect can generate
\begin{eqnarray}
m_u^\mathrm{eff} = 1.93 \pm 0.18 \ \mathrm{MeV},
\end{eqnarray}
when $m_u = 0$ in eq.(\ref{eq:massredefinition}).
This value differs from the value in eq.({\ref{eq:fittedmasses}) by about 4.7$\sigma$;
 $m_u = 0$ is not the solution of the strong CP problem.

Since $L_8$ does not seem to directly represent the effect of the 't~Hooft vertex,
 the coupling $L_8$ would not be produced dominantly by the instanton effect.
However, it is transformed under the instanton transformation,
 and therefore an evaluation of the maximum omega parameter is possible with $L_8$.
\footnote{
The quantitative investigation of the contribution of instanton effects to $L_8$
 is a future important task.
}
We obtain
\begin{eqnarray}
\label{eq:weaklyconstrained}
\omega_\mathrm{max} = 0.5 \pm 0.1.
\end{eqnarray}
Note that we do not need to neglect neutral meson mixing in this case which is necessary for $L_7$.
Since $L_8$ can have this contribution without instanton dynamics,
 eq.(\ref{eq:weaklyconstrained}) gives a ``weak constraint" on the maximum omega parameter.
The instanton-induced mass correction gives
\begin{eqnarray}
m_u^\mathrm{eff} = 2.33 \pm 0.20 \MeV
\end{eqnarray}
when $m_u = 0$.
This value differs from the value in eq.(\ref{eq:fittedmasses}) by about 2.3$\sigma$,
and $m_u = 0$ is not favored as a solution of the strong CP problem.
We have confirmed this known result in a simple way
 without loop effects of pseudo-Nambu-Goldstone bosons,
 which could be large contributions in chiral perturbation theory.
These results are consistent with the result $m_u/m_d \neq 0$ by Leutwyler in \cite{Leutwyler:1989pn},
 which supports the validity of our naive estimate.
These our results indicate that the instanton effect is small.

\section{Constraints in heavy meson system}
\label{sec:HMET}

In this section, we consider heavy meson systems using heavy meson effective theory
 which includes both light and heavy mesons \cite{Georgi:1991mr,Burdman:1992gh,Wise:1992hn}.
Since the instanton effect gives a correction of next-to-leading order in the expansion in ${\cal M}$,
 we construct the effective Lagrangian up to ${\cal O}(p^4)$ in the chiral expansion in heavy meson effective theory.

In heavy meson effective Lagrangian, heavy mesons are described by the effective field
\begin{eqnarray}
H_v(x) = \frac{ 1 + \Slash{v}}{2}[ P^\ast_{v\mu}(x) \gamma^\mu + i P_v(x) \gamma_5],
\end{eqnarray} 
where $v_\mu$ is the velocity of the heavy meson,
 and $H_v$ has mass dimension 3/2.
With two heavy flavors, $c$ and $b$, and three light flavors, $u$, $d$ and $s$,
the fields $P^\ast_v$ and $P_v$ include six heavy-light vector and six heavy-light pseudoscalar mesons
\begin{eqnarray}
P_v^{(\ast)} = \begin{pmatrix} D_u^{(\ast)} & D_d^{(\ast)} & D_s^{(\ast)} \\[2pt] B_u^{(\ast)} & B_d^{(\ast)} & B_s^{(\ast)} \end{pmatrix}.
\end{eqnarray}
The fields $B_u$, $B_d$ and $B_s$ correspond to the mesons $B^+$, $B^0$ and $B^0_s$, respectively.
The field $H_v$ is transformed under the spin-flavor SU(4) transformation,
 which can be decomposed by spin SU(2)$_\mathrm{s}$ and heavy flavor SU(2)$_\mathrm{f}$ transformations, as
\begin{eqnarray}
H_v &\rightarrow& SH_v, \\
H_v &\rightarrow& z_H H_v,
\end{eqnarray}
where $S$ $\in$ SU(2)$_\mathrm{s}$ act on the spinor index and
 $z_H$ $\in$ SU(2)$_\mathrm{f}$ act on the heavy flavor index.
The field is also transformed under the chiral SU(3)$_L\times$SU(3)$_R\times$U(1)$_A$ transformation as
\begin{eqnarray}
H_v &\rightarrow& H_v h(\Pi,V_L,V_R)^\dagger.
\end{eqnarray}
The parity transformation and charge conjugation are defined as follows,
\begin{eqnarray}
{\cal P} H_v(x) {\cal P}^\dagger &=& \gamma^0 H_{v_P} (x_P) \gamma^0, \\
{\cal C} H_v(x) {\cal C}^\dagger &=& C (\bar{H}_{-v}(x))^T C^\dagger,
\end{eqnarray}
where $v_P = (v^0, -{\bf v})$, $x_P = (x^0, -{\bf x})$ and $C=i\gamma^2\gamma^0$.
We need to introduce
 \begin{eqnarray}
 	\label{eq:Mdefinition}
 M(x) = \xi^\dagger(x) \exp\left({\frac{-i \eta'}{\sqrt{6}f'}}\right) \chi \xi^\dagger(x)\exp\left({\frac{-i \eta'}{\sqrt{6}f'}}\right)
\end{eqnarray}
to describe chiral symmetry breaking by the masses of light quarks
and it is transformed 
\begin{eqnarray}
M(x) \rightarrow h(\Pi,V_L,V_R) M(x) h(\Pi,V_L,V_R)^\dagger
\end{eqnarray}
under the chiral transformation, if light quark masses transform appropriately.
This is also transformed under the parity transformation and charge conjugation as
\begin{eqnarray}
{\cal P} M(x) {\cal P}^\dagger &=& M^\dagger(x_P), \\
{\cal C} M(x) {\cal C}^\dagger &=& M^T(x).
\end{eqnarray}

The effective Lagrangian can be obtained by imposing the chiral symmetry, spin-flavor symmetry,
 and invariance under the parity transformation, charge conjugation and Hermite conjugation.
 The effect of spin-flavor symmetry breaking is included by introducing the matrix
$1/M_Q \equiv$ diag$(1/M_c, 1/M_b)$,
 where $M_c$ and $M_b$ are charm and bottom quark masses, respectively.
Since we consider the situation that the heavy quark masses are much larger than the typical QCD scale,
 $\Lambda \equiv 4\pi f$, the effective Lagrangian is expanded in powers of $\Lambda/M_Q$.
We do not consider the terms which includes derivatives on $H_v$,
since we are going to analyse only the masses of heavy mesons.
We write down non-derivative part of the effective Lagrangian 
 up to ${\cal O}(\Lambda/M_Q)$ in the $\Lambda/M_Q$ expansion
 and ${\cal O}((m_q / \Lambda)^2)$ in the chiral expansion
\begin{eqnarray}
	\label{eq:HMEL}
{\cal {L}}_v^\mathrm{mass} &=&\Lambda \Bra\Hvs\Ket + \kappa' \Lambda \Bra\tr[\bar{H}_v \frac{\Lambda}{M_Q} H_v] \Ket + \kappa \Lambda \Bra \tr [\bar{H}_v \frac{\Lambda}{M_Q}\sigma_{\rho\sigma} H_v \sigma^{\rho\sigma}] \Ket \nonumber \\[5pt]
&&+ \frac{\chi_1}{\Lambda} \Bra \Hvs \Ket \Bra M + M^\dagger \Ket + \frac{\chi_2}{\Lambda} \Bra \Hvs (M + M^\dagger) \Ket \nonumber \\[5pt]
&&\quad+ \frac{a_1}{\Lambda} \Bra \tr[\bar{H}_v \frac{\Lambda}{M_Q} H_v] \Ket \Bra M + M^\dagger \Ket + \frac{a_2}{\Lambda} \Bra \tr [\bar{H}_v \frac{\Lambda}{M_Q} H_v] ( M + M^\dagger ) \Ket \nonumber \\[5pt] 
&&\quad+ \frac{b_1}{\Lambda} \Bra \tr[ \bar{H}_v \frac{\Lambda}{M_Q} \sigma_{\rho\sigma} H_v \sigma^{\rho\sigma} ]\Ket \Bra M + M^\dagger \Ket + \frac{b_2}{\Lambda} \Bra \tr[ \bar{H}_v \frac{\Lambda}{M_Q} \sigma_{\rho\sigma} H_v \sigma^{\rho\sigma} ]( M + M^\dagger ) \Ket \nonumber \\[5pt]
&&\quad + \frac{K_1}{\Lambda^3} \Bra \Hvs \Ket \Bra M \Ket \Bra M^\dagger \Ket \nonumber \\[5pt]
&&\quad + \frac{K_2}{\Lambda^3} \left\{ \Bra \Hvs M \Ket \Bra M^\dagger \Ket + \Bra \Hvs M^\dagger \Ket \Bra M \Ket \right\} \nonumber \\[5pt]
&&\quad + \frac{K_3}{\Lambda^3} \left\{ \Bra \Hvs MM^\dagger \Ket + \Bra \Hvs M^\dagger M \Ket \right\} \nonumber \\[5pt]
&&\quad + \frac{K_4}{\Lambda^3} \Bra \Hvs \Ket \left\{ \Bra MM  +  M^\dagger M^\dagger \Ket \right\} \nonumber \\[5pt]
&&\quad + \frac{K_5}{\Lambda^3} \Bra \Hvs \Ket \left\{ \Bra M \Ket \Bra M \Ket + \Bra M^\dagger \Ket \Bra M^\dagger \Ket \right\} \nonumber \\[5pt]
&&\quad + \frac{K_6}{\Lambda^3} \left\{ \Bra \Hvs M \Ket \Bra M \Ket + \Bra \Hvs M^\dagger \Ket \Bra M^\dagger \Ket \right\} \nonumber \\[5pt]
&&\quad + \frac{K_7}{\Lambda^3} \left\{ \Bra \Hvs MM \Ket + \Bra \Hvs M^\dagger M^\dagger \Ket \right\} \nonumber \\
&+& {\cal{O}}\left( \left( \frac{\Lambda}{M_Q} \right)^2 \right),
\end{eqnarray}
where $\sigma^{\rho\sigma} = i[\gamma^\rho,\gamma^\sigma]/2$.
The couplings $\kappa'$, $\kappa$, $\chi_i$, $a_i$, $b_i$ and the seven couplings $K_i$ are dimensionless parameters.
A possible term
\begin{eqnarray}
\frac{K_0}{\Lambda} \Bra \Hvs \Ket \Bra M^\dagger M + M M^\dagger \Ket
\end{eqnarray}
 can be absorbed into the first term in ${\cal L}_v^\mathrm{mass}$ due to the unitarity of $\xi$.
The terms with coupling $K_i$ gives a complete collection of terms of ${\cal O}(p^4)$ without derivatives,
though some of the terms have already given by \cite{Jenkins:1992hx}.
The order $(\Lambda/M_Q)^2$ terms,
which could give the contribution of the same order of ${\cal O}(p^4)$ terms,
 are irrelevant to our present analysis,
since we are not going to consider the mass differences between heavy pseudoscalar mesons and heavy vector mesons.

The heavy meson effective Lagrangian has also the invariance
 under the instanton transformation of eq.(\ref{eq:instantontransformation}) with
\begin{equation}
\label{eq:Kstransformation}
	\begin{split}
K_4 &\rightarrow K_4 + \frac{\chi_1 + \chi_2}{2} \omega, \\
K_5 &\rightarrow K_5 - \frac{\chi_1 + \chi_2}{2} \omega, \\
K_6 &\rightarrow K_6 + \chi_2 \omega, \\
K_7 &\rightarrow K_7 -  \chi_2 \omega.
	\end{split}
\end{equation}
This invariance can be shown by using eq.(\ref{eq:matrixidentity}).
The terms with couplings $K_4$, $K_5$, $K_6$ and $K_7$ break U(1)$_A$ symmetry,
 because $M$ is transformed under U(1)$_A$ axial transformation 
 (note that in eq.(\ref{eq:Mdefinition}) $\eta'$ transforms as $\eta' \rightarrow \eta' + \sqrt{6} f' \theta_A /2$
 and $\chi$ is invariant since it is essentially the mass matrix).
Remember that the instanton effect breaks U(1)$_A$ symmetry,
and in fact, these couplings are sensitive to the instanton transformation.
On the other hand, the other terms of $K_1$, $K_2$ and $K_3$ are invariant under U(1)$_A$ axial transformation,
 and insensitive to the instanton transformation.

In the following, we fit a combination of couplings, $K_3 + K_7$, which is sensitive to the instanton correction
 using the well-known masses of pseudoscalar B mesons only.
 
We obtain pseudoscalar B meson mass formulae from eq.(\ref{eq:HMEL})
\begin{eqnarray}
\label{eq:heavymassformulae}
M^2_{B_q} =  M_b^2 \Biggl[ 1 &+& \frac{\Lambda}{M_b} \Biggl\{ 2 + 2 \left( \kappa' + 6 \kappa \right) \frac{\Lambda}{M_b} + 8\left( \chi_1 +  a_1 \frac{\Lambda}{M_b} + 6 b_1 \frac{\Lambda}{M_b} \right) \frac{B_0}{\Lambda} \frac{m_u + m_d + m_s}{\Lambda} \nonumber \\
&& \quad + 8 \left( K_1 + 2 K_5 \right) \frac{B_0^2}{\Lambda^2} \frac{(m_u + m_d + m_s)^2}{\Lambda^2} + 16 K_4  \frac{B_0^2}{\Lambda^2} \frac{m_u^2 + m_d^2 + m_s^2}{\Lambda^2} \Biggr\} \nonumber \\
&+&  \frac{\Lambda}{M_b} \Biggl\{ 8 \left( \chi_2 + a_2 \frac{\Lambda}{M_b} + 6 b_2 \frac{\Lambda}{M_b} \right) \frac{B_0}{\Lambda} + 16 \left(K_2 + K_6 \right) \frac{B_0^2}{\Lambda^2} \frac{m_u + m_d + m_s}{\Lambda} \Biggr\}\frac{m_q}{\Lambda} \nonumber \\
&+& \frac{\Lambda}{M_b}16\left( K_3 + K_7 \right) \frac{B_0^2}{\Lambda^2} \frac{m_q^2}{\Lambda^2} \Biggl] + {\cal O}\left(\left(\frac{\Lambda}{M_Q}\right)^2\right),
\end{eqnarray}
where $q$ is the light flavor index, $u,d$ and $s$.
In this heavy meson mass formulae,
 we see that the couplings $K_5$ and $K_6$ describe direct contribution of the instanton-induced mass
 like Fig.\ref{fig:instantoncorr},
 though the couplings $K_4$ and $K_7$ should be also sensitive to the instanton transformation.
This is the same argument on $L_6$, $L_7$ and $L_8$ in previous section.
We can fit
\begin{eqnarray}
\label{eq:chi_2}
\chi_2 = \frac{\Lambda}{4} \frac{\hat{M}_{B_s} - \hat{M}_{B_d}}{\hat{m}^2_{K^\pm} - \hat{m}^2_{\pi^\pm}}=0.065 \pm 0.004,
\end{eqnarray}
at ${\cal O}(p^2)$.
Up to ${\cal O}(p^4)$ the meson mass differences are obtained as follows
\begin{eqnarray}
M_{B_s} - M_{B_d} &=&\frac{B_0(m_s - m_d)}{\Lambda} \Biggl[ 4 \chi_2 + (4a_2 + 24 b_2) \frac{\Lambda}{M_b} + 8 (K_2 + K_6) \frac{B_0(m_u + m_d + m_s)}{\Lambda^2} \nonumber \\
&& \qquad \qquad \qquad \qquad \qquad \qquad \qquad \qquad \qquad + 8(K_3 + K_7) \frac{B_0(m_s + m_d)}{\Lambda^2} \Biggr], \\
M_{B_d} - M_{B_u} &=&\frac{B_0(m_d - m_u)}{\Lambda} \Biggl[ 4 \chi_2 + (4a_2 + 24 b_2) \frac{\Lambda}{M_b} + 8 (K_2 + K_6) \frac{B_0(m_u + m_d + m_s)}{\Lambda^2} \nonumber \\
&& \qquad \qquad \qquad \qquad \qquad \qquad \qquad \qquad \qquad + 8(K_3 + K_7) \frac{B_0(m_d + m_u)}{\Lambda^2} \Biggr].
\end{eqnarray}
Therefore, we can extract only one combination
\begin{eqnarray}
K_3 + K_7 &=& \frac{\Lambda^3}{8(\hat{m}^2_{K^0} - \hat{m}^2_{\pi^0})} \Biggl\{ \frac{\hat{M}_{B_s} - \hat{M}_{B_d}}{\hat{m}^2_{K^\pm} - \hat{m}^2_{\pi^\pm}} - \frac{\hat{M}_{B_d} - \hat{M}_{B_u} + (\hat{M}_{B_u} - \hat{M}_{B_d})_\mathrm{EM}}{\hat{m}^2_{K^0} - \hat{m}^2_{K^\pm} + \hat{m}^2_{\pi^\pm} - \hat{m}^2_{\pi^0}} \Biggr\} \nonumber \\
&& - \frac{\chi_2}{2} \frac{\Lambda^2}{\hat{m}^2_{K^\pm} + \hat{m}^2_{K^0} - \hat{m}^2_{\pi^\pm}} \Biggl\{ \frac{\hat{m}^2_{K^0} - \hat{m}^2_{K^\pm} + \hat{m}^2_{\pi^\pm} - \hat{m}^2_{\pi^0}}{\hat{m}^2_{\pi^0}} \frac{m_d + m_u}{m_d - m_u} -1 \Biggr\},
\end{eqnarray}
where we have considered the QED correction to the masses of the charged mesons.
Substituting 
\begin{eqnarray}
\label{eq:EMcorrection}
(M_{B_u} - M_{B_d})_\mathrm{EM} = 2.09 \pm 0.18 \MeV,
\end{eqnarray}
which is given by a theoretical calculation in \cite{Goity:2007fu},
 and the observed value of the mass differences in \cite{Olive:2016xmw},
 we obtain 
\begin{eqnarray}
\label{eq:Kvalue}
K_3 + K_7 = -0.013 \pm 0.007
\end{eqnarray}

Since $K_7$ does not seem to describe direct contribution of the instanton effect in mass formulae of eq.(\ref{eq:heavymassformulae}),
 and $K_3$ does not transform under the instanton transformation,
 we are going to argue ``weak constraint" on the possible maximal value of the parameter $\omega$.
This should be equivalent what we have obtained from $L_8$ in the previous section.
\footnote{
The quantitative investigation of the contribution of instanton effects to $K_7$
 is a future important task.
We believe that the values of $K_3$ and $K_7$ will be independently extracted
 from the data by future experiments.
}
It can be obtained as
\begin{eqnarray}
\omega_\mathrm{max} = 0.2 \pm 0.1.
\end{eqnarray}
 by using eqs.(\ref{eq:Kstransformation}) and (\ref{eq:chi_2})
 and by assuming that all of the value in eq.(\ref{eq:Kvalue}) is produced by the instanton dynamics.

A constraint on the possible maximal value of the omega parameter can be also evaluated {\it in heavy meson system}.
It is found that the constraint on the instanton effect in heavy meson system is as strong as
 that in the light meson system.
In this case, the instanton-induced effective up-quark mass is given as
\begin{eqnarray}
m_u^\mathrm{eff} = 1.06 \pm 0.50 \ \MeV
\end{eqnarray}
when $m_u = 0$ by eq.(\ref{eq:massredefinition}).
This value is inconsistent with that in eq.(\ref{eq:fittedmasses})
 and the solution of strong CP problem by $m_u = 0$ is excluded.
 Note that, if the sign of non-instanton $K_3$ is opposite of the sign of $K_7$,
 we are overconstraining the instanton effect.
To avoid this, we must determine the value of $K_3$, which should be future work.
Also note that, there would be room for improvement to take some possible loop effects
 of pseudo-Nambu-Goldstone bosons into account
 as we have noted in previous section.
Our analysis using the heavy meson effective theory indicates
 that the instanton-induced mass correction does not seem to be large enough to solve the strong CP problem.

\section{Conclusions}
\label{sec:concl}

We have investigated the light meson effective Lagrangian
 of the system with pseudo Nambu-Goldstone bosons at low energies up to ${\cal O}(p^4)$,
 and have confirmed that the effective Lagrangian is invariant under the instanton transformation.
This symmetry is the consequence of the fact that the physics is independent 
from whether the instanton correction is included in $\chi$ or $L_6,$ $L_7$ and $L_8$.
We have evaluated the value of the coupling $L_7$,
 which is expected to dominantly produced by the instanton dynamics.
The maximum value of the omega parameter was calculated
 under the assumption that whole the value of $L_7$ is produced by the instanton dynamics.
The result $\omega_\mathrm{max} = 0.8 \pm 0.1$ is not able to produce the situation
 that $m_u = 0$ could be a solution of strong CP problem.
In other words, instanton corrections to quark masses are small.
The ``weakly constrained" maximum omega parameter has been also calculated with the coupling $L_8$
 though the coupling might have the contribution from the other non-instanton dynamics in QCD.
The result in this case also indicates that the instanton effect is small.
We have confirmed this known result in a simple way.

We have shown that the same analysis is possible for heavy meson systems.
We have constructed the heavy meson effective Lagrangian with chiral symmetry breaking
 up to ${\cal O}(p^4)$ in the chiral expansion.
This effective Lagrangian also is invariant under the instanton transformation
 and we have identified the couplings which are non-trivially transformed under the instanton transformation.
Bottomed pseudoscalar meson mass formulae up to ${\cal O}(p^4)$ have been derived,
 since their masses have already been measured well by experiments
 and $\Lambda/M_b$ expansion is more reliable than $\Lambda/M_c$ expansion in heavy meson effective theory.
Although we could only determine a combination of $K_3 + K_7$
 ($K_3$ is insensitive and  $K_7$ is sensitive to the instanton transformation),
 the possible maximal value (``weak constraint") $\omega_\mathrm{max} = 0.2 \pm 0.1$
 is obtained from the meson mass differences
 under the assumption that $K_3+K_7$ is dominated by instanton effects,
 which should be quantitatively investigated in future.
It has been found that the constraint on the instanton parameter in heavy meson system
 is as strong as that in the light meson system.
The development of lattice calculations is going to give a more precise omega parameter,
since the error of omega parameter is mainly from the error of quark condensation given by lattice calculations.

Our results are obtained following the procedure based on a spirit of chiral expansion theory,
 where the parameters are fitted order by order.
When the light quark masses are fitted in leading order,
 the freedom of instaton transformation is fixed.
The instanton corrections are included in higher order terms.
The instanton correction can be absorbed into leading order quark masses with a special instanton transformation
 by which higher order couplings vanish under a conservative assumption of the dominance of instanton contributions
 to the higher order couplings.
The value of $\omega_\mathrm{max}$ corresponds to such an instanton transformation,
 and it can be a measure of the magnitude of instanton effect as well as $m_u^{\mathrm{eff}}$.

In the following, we point out the problems and future subjects of our present analysis in order.

We note that both the results in light and heavy meson systems were obtained under the assumption
 that whole the values of the corresponding couplings were generated by the instanton dynamics only. 
If the contribution of non-instanton dynamics in QCD to the couplings is large,
 especially it cancels the contribution of instanton dynamics,
 our constraints do not apply.

In order to make the constraint more precise for heavy meson systems,
 we must determine the value of $K_3$ independently.
In the future, the information of $b$-flavored vector meson masses given by experiments
 will enable us to fit more precisely and systematically the couplings $K_i$.
When we consider the mass differences between the vector mesons and pseudoscalar mesons,
 we need to include the ${\cal O}((\Lambda/M_b)^2)$ terms in the heavy meson effective Lagrangian.

As the first step of the analysis, we have neglected, for simplicity, the chiral symmetry breaking terms in leading order with derivative of $H_v$,
$\tr(H v\cdot \di H)\bra M \ket$ and $\bra \tr(H v\cdot \di H) M \ket$,
 which require field redefinitions, and mass formulae should obtain corrections.
Again, we hope that the results of future B factory experiments enable us to include these terms.

We also have neglected the loop effects of pseudo-Nambu-Goldstone bosons in chiral perturbation theory.
We leave this problem to future work.

According to \cite{Goity:2007fu}, the masses of $B_d$ and $B_u$ are almost degenerate
 as a result of the cancellation of two sources of isospin breaking: mass difference of up- and down-quarks and QED effect.
This situation is very different from that in light mesons or charmed mesons, $\pi$, $K$ and $D$.
The phenomena should be studied better to obtain more precise input of the theoretical QED effect.

In this article a new quantitative constraint to the instanton effect in heavy meson system has been given.
We hope that future development of this approach will help to verify the instanton effect experimentally.

\section*{Acknowledgements}
We would like to thank S.~Fukasawa and K.~Masukawa for helpful comments and discussions.
N.K. was supported in part by JSPS KAKENHI Grant Number 26400253.
Y.S. was supported in part by JGC-S SCHOLARSHIP FOUNDATION.


\end{document}